\input harvmac
\overfullrule=0pt
\parindent 25pt
\tolerance=10000

\input epsf

\newcount\figno
\figno=0
\def\fig#1#2#3{
\par\begingroup\parindent=0pt\leftskip=1cm\rightskip=1cm\parindent=0pt
\baselineskip=11pt
\global\advance\figno by 1
\midinsert
\epsfxsize=#3
\centerline{\epsfbox{#2}}
\vskip 12pt
{\bf Fig.\ \the\figno: } #1\par
\endinsert\endgroup\par
}
\def\figlabel#1{\xdef#1{\the\figno}}
\def\encadremath#1{\vbox{\hrule\hbox{\vrule\kern8pt\vbox{\kern8pt
\hbox{$\displaystyle #1$}\kern8pt}
\kern8pt\vrule}\hrule}}

 \def\ep{{\epsilon}}

 \def\frac#1#2{{#1\over #2}}

 \def\s{\sqrt}

 \def\al{\alpha'}
 \def\de{\partial}

 \def\lr{\leftrightarrow}
 \def\f {\frac}
 \def\ti{\tilde}
 \def\ap{\alpha}

 \def\lb{\rangle}
 \def\ep{\epsilon}

 \def\vp{\varphi}

\lref\GRa{
D.~Gaiotto and L.~Rastelli,
``A paradigm of open/closed duality: Liouville D-branes and the Kontsevich
model,''
arXiv:hep-th/0312196.
}

\lref\Kaz{
V.~A.~Kazakov,
``A Simple Solvable Model Of Quantum Field Theory Of Open Strings,''
Phys.\ Lett.\ B {\bf 237}, 212 (1990).
}

\lref\MoSe{
G.~W.~Moore and N.~Seiberg,
``From loops to fields in 2-D quantum gravity,''
Int.\ J.\ Mod.\ Phys.\ A {\bf 7}, 2601 (1992).
}

\lref\Se{
G.~W.~Moore, N.~Seiberg and M.~Staudacher,
``From loops to states in 2-D quantum gravity,''
Nucl.\ Phys.\ B {\bf 362}, 665 (1991).
}

\lref\GTT{
S.~Gukov, T.~Takayanagi and N.~Toumbas,
``Flux backgrounds in 2D string theory,''
arXiv:hep-th/0312208.
}

\lref\FZZ{
V.~Fateev, A.~B.~Zamolodchikov and A.~B.~Zamolodchikov,
``Boundary Liouville field theory. I: Boundary state and boundary
two-point
function,''
arXiv:hep-th/0001012.
}

\lref\TB{J.~Teschner,
``Liouville theory revisited,''
Class.\ Quant.\ Grav.\  {\bf 18}, R153 (2001)
[arXiv:hep-th/0104158].
}

\lref\Bo{
D.~V.~Boulatov,
``Critical Behavior In The Model Of Random Surfaces With Dynamical Holes
Embedded In Two-Dimensional And One-Dimensional Spaces,''
Int.\ J.\ Mod.\ Phys.\ A {\bf 6}, 79 (1991).
}

\lref\Ya{
Z.~Yang,
``Dynamical Loops In D = 1 Random Matrix Models,''
Phys.\ Lett.\ B {\bf 257}, 40 (1991).
}

\lref\MV{
J.~McGreevy and H.~Verlinde,
``Strings from tachyons: The c = 1 matrix reloaded,''
[arXiv:hep-th/0304224].}

\lref\KMS{
I.~R.~Klebanov, J.~Maldacena and N.~Seiberg,
``D-brane decay in two-dimensional string theory,''
[arXiv:hep-th/0305159].}

\lref\MTV{
J.~McGreevy, J.~Teschner and H.~Verlinde,
``Classical and quantum D-branes in 2D string theory,''
[arXiv:hep-th/0305194].
}

\lref\TT{T.~Takayanagi and N.~Toumbas,
``A matrix model dual of type 0B string theory in two dimensions,''
JHEP {\bf 0307}, 064 (2003)
[arXiv:hep-th/0307083].
}

\lref\KKK{V.~Kazakov, I.~K.~Kostov and D.~Kutasov,
``A matrix model for the two-dimensional black hole,''
Nucl.\ Phys.\ B {\bf 622}, 141 (2002)
[arXiv:hep-th/0101011].
}

\lref\KlR{I.~R.~Klebanov,
``String theory in two-dimensions,''
[arXiv:hep-th/9108019].}

\lref\six{M.~R.~Douglas, I.~R.~Klebanov, D.~Kutasov, J.~Maldacena,
E.~Martinec and N.~Seiberg,
``A new hat for the c = 1 matrix model,''
[arXiv:hep-th/0307195].
}

\lref\FuHo{
T.~Fukuda and K.~Hosomichi,
``Super Liouville theory with boundary,''
Nucl.\ Phys.\ B {\bf 635}, 215 (2002)
[arXiv:hep-th/0202032].
}

\lref\JY{A.~Jevicki and T.~Yoneya,
``A Deformed matrix model and the black hole background
in two-dimensional string theory,''
Nucl.\ Phys.\ B {\bf 411}, 64 (1994)
[arXiv:hep-th/9305109].
}

\lref\poc{J.~Polchinski,
``Classical Limit Of (1+1)-Dimensional String Theory,''
Nucl.\ Phys.\ B {\bf 362}, 125 (1991).
}

\lref\DJ{S.~R.~Das and A.~Jevicki,
``String Field Theory And Physical Interpretation Of D = 1 Strings,''
Mod.\ Phys.\ Lett.\ A {\bf 5}, 1639 (1990).
}

\lref\wittenbh{E.~Witten,
``On string theory and black holes,''
Phys.\ Rev.\ D {\bf 44}, 314 (1991).
}

\lref\poreview{J.~Polchinski,
``What is string theory?,''
[arXiv:hep-th/9411028].
}

\lref\ponon{
J.~Polchinski,
``Combinatorics Of Boundaries In String Theory,''
Phys.\ Rev.\ D {\bf 50}, 6041 (1994)
[arXiv:hep-th/9407031].
}

\lref\AKKT{
S.~Y.~Alexandrov, V.~A.~Kazakov and I.~K.~Kostov,
``Time-dependent backgrounds of 2D string theory,''
Nucl.\ Phys.\ B {\bf 640}, 119 (2002)
[arXiv:hep-th/0205079].
}

\lref\GM{
P.~Ginsparg and G.~W.~Moore,
``Lectures On 2-D Gravity And 2-D String Theory,''
[arXiv:hep-th/9304011].
}

\lref\ARS{
C.~Ahn, C.~Rim and M.~Stanishkov,
``Exact one-point function of N = 1 super-Liouville theory with boundary,''
Nucl.\ Phys.\ B {\bf 636}, 497 (2002)
[arXiv:hep-th/0202043].
}

\lref\ads{J.~M.~Maldacena,
``The large N limit of superconformal field theories and supergravity,''
Adv.\ Theor.\ Math.\ Phys.\  {\bf 2}, 231 (1998)
[Int.\ J.\ Theor.\ Phys.\  {\bf 38}, 1113 (1999)]
[arXiv:hep-th/9711200].
}

\lref\Ma{E.~J.~Martinec,
``The annular report on non-critical string theory,''
[arXiv:hep-th/0305148].
}
\lref\GK{
M.~Gutperle and P.~Kraus,
``D-brane dynamics in the c = 1 matrix model,''
[arXiv:hep-th/0308047].
}
\lref\Sen{
A.~Sen,
``Open-closed duality: Lessons from matrix model,''
[arXiv:hep-th/0308068].
}

\lref\MMV{
J.~McGreevy, S.~Murthy and H.~Verlinde,
``Two-dimensional superstrings and the supersymmetric matrix model,''
arXiv:hep-th/0308105.
}

\lref\Ka{A.~Kapustin,
``Noncritical superstrings in a Ramond-Ramond background,''
[arXiv:hep-th/0308119].
}
\lref\KaSt{
J.~L.~Karczmarek and A.~Strominger,
``Matrix cosmology,''
[arXiv:hep-th/0309138].
}
\lref\DSVY{
J.~de Boer, A. ~Sinkovics, E. ~Verlinde, J.T. ~Yee,
``String Interactions in c=1 Matrix Model,''
[arXiv:hep-th/0312135].
}
\lref\Janik{J.~Ambjorn, R.~A.~Janik, ``The decay of quantum D-branes,''
[arXiv:hep-th/0312163].
}
\lref\Sh{N.~Seiberg, D.~Shih, ``Branes, Rings and
Matrix Models in Minimal (Super)string Theory,'' [arXiv:hep-th/0312170].
}

\lref\Nakayama{
Y.~Nakayama,
``Liouville field theory: A decade after the revolution,''
arXiv:hep-th/0402009.
}
\lref\BSFT{
E.~Witten,
``On background independent open string field theory,''
Phys.\ Rev.\ D {\bf 46}, 5467 (1992)
[arXiv:hep-th/9208027];
A.~A.~Gerasimov and S.~L.~Shatashvili,
``On exact tachyon potential in open string field theory,''
JHEP {\bf 0010}, 034 (2000)
[arXiv:hep-th/0009103];
D.~Kutasov, M.~Marino and G.~W.~Moore,
``Some exact results on tachyon condensation in string field theory,''
JHEP {\bf 0010}, 045 (2000)
[arXiv:hep-th/0009148].
}

\lref\BSFTS{ D.~Kutasov, M.~Marino and G.~W.~Moore, ``Remarks on
tachyon condensation in superstring field theory,''
arXiv:hep-th/0010108.
}

\lref\BSFTBA{
P.~Kraus and F.~Larsen,
``Boundary string field theory of the DD-bar system,''
Phys.\ Rev.\ D {\bf 63}, 106004 (2001)
[arXiv:hep-th/0012198];
T.~Takayanagi, S.~Terashima and T.~Uesugi,
``Brane-antibrane action from boundary string field theory,''
JHEP {\bf 0103}, 019 (2001)
[arXiv:hep-th/0012210].
}

\lref\RT{
A.~Sen,
``Rolling tachyon,''
JHEP {\bf 0204}, 048 (2002)
[arXiv:hep-th/0203211].
}

\lref\SenR{ A.~Sen, ``Rolling tachyon boundary state, Conserved
charges and Two dimensional string theory''
 [arXiv:hep-th/0402157].
}

\lref\RTS{
A.~Sen,
``Tachyon matter,''
JHEP {\bf 0207}, 065 (2002)
[arXiv:hep-th/0203265].
}

\lref\GSSB{
M.~Gutperle and A.~Strominger,
``Spacelike branes,''
JHEP {\bf 0204}, 018 (2002)
[arXiv:hep-th/0202210].
}

\lref\Das{
S.~R.~Das,
``D branes in 2d string theory and classical limits,''
arXiv:hep-th/0401067.
}

\lref\Ko{
I.~K.~Kostov,
``Boundary ground ring in 2D string theory,''
arXiv:hep-th/0312301.
}

\lref\BKR{
A.~Boyarsky, B.~Kulik and O.~Ruchayskiy,
``Classical and quantum branes in c = 1 string theory and quantum Hall
arXiv:hep-th/0312242.
}

\lref\MW{
G.~Mandal and S.~R.~Wadia,
``Rolling tachyon solution of two-dimensional string theory,''
arXiv:hep-th/0312192.
}

\lref\AJ{
J.~Ambjorn and R.~A.~Janik,
``The decay of quantum D-branes,''
arXiv:hep-th/0312163.
}

\lref\ADKMV{
M.~Aganagic, R.~Dijkgraaf, A.~Klemm, M.~Marino and C.~Vafa,
``Topological strings and integrable hierarchies,''
arXiv:hep-th/0312085.
}

\lref\ES{
T.~Eguchi and Y.~Sugawara,
``Modular bootstrap for boundary N = 2 Liouville theory,''
JHEP {\bf 0401}, 025 (2004)
[arXiv:hep-th/0311141].
}

\lref\Muk{
S.~Mukhi,
``Topological matrix models, Liouville matrix model
and c = 1 string theory,''
arXiv:hep-th/0310287.
}

\lref\Ale{
S.~Alexandrov,
``(m,n) ZZ branes and the c = 1 matrix model,''
arXiv:hep-th/0310135.
}

\lref\DaDa{
S.~Dasgupta and T.~Dasgupta,
``Renormalization group approach to c = 1 matrix model
 on a circle and D-brane
decay,''
arXiv:hep-th/0310106.
}

\lref\RSc{
S.~Ribault and V.~Schomerus,
``Branes in the 2-D black hole,''
arXiv:hep-th/0310024.
}

\lref\Tesr{
J.~Teschner,
``On boundary perturbations in Liouville theory and brane dynamics in
noncritical string theories,''
arXiv:hep-th/0308140.
}

\lref\GIR{
D.~Gaiotto, N.~Itzhaki and L.~Rastelli,
``On the BCFT description of holes in the c = 1 matrix model,''
Phys.\ Lett.\ B {\bf 575}, 111 (2003)
[arXiv:hep-th/0307221].
}

\lref\KPS{
I.~K.~Kostov, B.~Ponsot and D.~Serban,
``Boundary Liouville theory and 2D quantum gravity,''
arXiv:hep-th/0307189.
}

\lref\AKK{
S.~Y.~Alexandrov, V.~A.~Kazakov and D.~Kutasov,
``Non-perturbative effects in matrix models and D-branes,''
JHEP {\bf 0309}, 057 (2003)
[arXiv:hep-th/0306177].
}

\lref\DKR{ K.~Demeterfi, I.~R.~Klebanov and J.~P.~Rodrigues, ``The
Exact S matrix of the deformed c = 1 matrix model,'' Phys.\ Rev.\
Lett.\  {\bf 71}, 3409 (1993) [arXiv:hep-th/9308036].
}

\baselineskip 18pt plus 2pt minus 2pt

\Title{\vbox{\baselineskip12pt
\hbox{hep-th/0402196}\hbox{HUTP-04/A010}
  }}
{\vbox{\centerline{Notes on D-branes in 2D Type 0
String Theory}}}
\centerline{Tadashi Takayanagi\foot{e-mail:
takayana@bose.harvard.edu}}

\medskip\centerline{ Jefferson Physical Laboratory}
\centerline{Harvard University}
\centerline{Cambridge, MA 02138, USA}

\vskip .1in \centerline{\bf Abstract} In this paper we construct
complete macroscopic operators in two dimensional type 0 string
theory. They represent D-branes localized in the time direction.
We give another equivalent description of them as deformed Fermi
surfaces. We also discuss a continuous array of such D-branes and
show that it can be described by a matrix model with a deformed
potential. For appropriate values of parameters, we find that it
has an additional new sector hidden inside its strongly coupled
region.

\noblackbox

\Date{February 2004}

\writetoc

\newsec{Introduction}

The two dimensional string theory (see e.g. \refs{\KlR, \GM,
\poreview}\ for reviews) is a very instructive model when we would
like to understand the nature of string theory as a complete
theory of quantum gravity. This theory has a powerful dual
description of $c=1$ matrix model defined by the simple quantum
mechanics of a Hermitian matrix $\Phi$ with the inverse harmonic
potential $U(\Phi)=-\Phi^2$ after the double scaling limit. In
particular, the matrix model dual of the two dimensional type 0
string \refs{\TT, \six}\ gives a non-perturbatively well-defined
formulation. This was constructed by employing a recent remarkable
interpretation of $c=1$ matrix model as a theory of multiple
unstable D0-branes \refs{\MV, \KMS, \MTV}. For example, the type
0B model is defined by the hermitian matrix model with two Fermi
surfaces \refs{\TT, \six}.

We expect this formulation will also offer us important clues to
understand the non-linear and non-perturbative backreactions in
quantum gravity when we put a macroscopic system like a large
number of D-branes. Note also that in the two dimensional string
theory we have no supersymmetry and thus may have a chance to
understand properties of (non-extremal) black holes in quantum
gravity. Motivated by this, in this paper we consider a particular
class of D-branes which are described by the macroscopic operators
 in the matrix model\foot{ See also the recent papers
\refs{\Sh, \GRa}\ for the discussions of macroscopic operators in
the $c<1$ matrix models.} (in the bosonic string context see
\refs{\MoSe, \Se, \Ma}; also refer to \refs{\Kaz, \Bo, \Ya}\ for
the similar methods of putting boundaries). They are those
D-branes\foot{ For recent discussions on other aspects of D-branes
in two dimensional string theory see e.g. \refs{\AKK, \KPS, \GIR,
\GK, \Sen, \MMV, \Tesr, \RSc, \DaDa, \Ale, \Muk, \ES, \DSVY, \AJ, \MW,
\BKR, \Ko, \Das, \Nakayama, \SenR}.}
 which are localized in the time direction $x^0$
and which extends along the Liouville direction
$\phi$ \refs{\FZZ, \TB, \FuHo, \ARS}.
As we will see later, we can indeed construct the corresponding operators in
type 0 string such that they reproduce the results
of boundary states \refs{\FuHo,\ARS}
perfectly. The backreactions due to the presence of these D-branes can be
conveniently described by the deformation of the Fermi surface\foot{
Refer to \refs{\AKKT,\KaSt}\
for general aspects of time-dependent Fermi surfaces in matrix model.}.
We can also
consider a continuous array of such D-branes in the time direction by taking
Fourier transformation. In this case we can show that the potential of the
matrix model will be deformed. In some cases we find an interesting
phenomenon that an additional new sector appears in the strongly couple region
due to the backreaction in the presence of the branes.

This paper is organized as follows. In section 2 we construct the
macroscopic operators in type 0 string and compare it with the
computations of boundary states. We also discuss an interpretation
of the operators by using the boundary string field theory. In
section 3 we consider how the matrix model will be deformed by the
presence of the D-branes. In section 4 we summarize and discuss
the results obtained.

\newsec{Macroscopic Operators and D-brane Boundary States}

The macroscopic loop operator $W(t,l)$ \refs{\MoSe, \Se}\ is such
an operator that creates a macroscopic hole with the length $l$ on
the world-sheet of non-critical string theory at time $t$. This is
roughly described by \eqn\mac{W(t,l)\sim
\delta\left(\int_{\de\Sigma}e^{\phi}-l\right)\cdot \delta(X^0-t).}

In the bosonic string it was identified with
$W_{bos}(t,l)=e^{-l\Phi(t)}$ \refs{\MoSe, \Se}. In fact, this
produces a correct vertex in the matrix model, which can be
regarded as an expected hole on the discretized world-sheet in the
$c=1$ matrix model (see e.g. the review \GM). The boundary length
in non-critical string corresponds to $e^{\phi}$ in terms of the
Liouville field.
 The physical meaning of this operator in two
dimensional string theory is the presence of a `Euclidean D-brane'
(localized in the time direction) \FZZ \Ma. To be more precise
after we take the Laplace transformation $\int d\phi e^{-\mu_B
e^{\phi}}$, we get a D-brane with the Neumann boundary condition
 in the Liouville direction (FZZT-brane \refs{\FZZ, \TB})
and the Dirichlet one in the time direction \eqn\fzztb{\int
\f{dl}{l}e^{-\mu_B l}\ W_{bos}(t,l) \simeq
|B_{(FZZT)}(\mu_B)\lb_{\phi}\otimes |D\lb_{X^0}.} The parameter
$\mu_B$ corresponds to the boundary cosmological constant in the
boundary state.
 Indeed we can show this relation \fzztb\ by
computing one point function on the brane or equally annulus
amplitude as shown in \Ma. However, as is obvious from our
later arguments, for negative values of $\mu_B$ or for large
number of branes, we expect non-perturbative effects become
important. To go beyond this problem, we need a non-perturbative
formulation.

Macroscopic operators in type 0 theory should also be defined as a
natural generalization of that in bosonic string. Since the two
dimensional type 0 string  is non-perturbatively stable, we can
expect that the operators are meaningful when we consider
non-perturbative corrections. In \TT\ one proposal was given such
that it respects the $Z_2$ symmetry $\Phi\to -\Phi$ of the open
string tachyon field and that it gives the correct leg-factor. In
this section we would like to give a complete set of macroscopic
operators extending the previous results in \TT\ so that it
explains the result of boundary states in super-Liouville theory
perfectly.

\subsec{Macroscopic Operators in Type 0 Matrix Model}

The macroscopic operators can be divided into NSNS and RR sector
part such that they correspond to the NSNS and RR sector part of
the D-brane boundary state. Moreover, since we know that there are
two types\foot{Notice that we do not impose the chiral GSO
projection in type 0 theory. The two different boundary states
survive the non-chiral projection.} of (FZZT-like) boundary states
$|B(\ep)\lb$ according to the spin structures \FuHo, there should
be two macroscopic operators $W^{(\ep)}$ with $\ep=\pm$.

We would like to argue that they are given by\foot{ The structure
of the operators shown here might suggest an existence of a more
fundamental definition in a certain superspace. The interpretation
by boundary superstring field theory discussed later may give a
hint about this point.} (at fixed energy $E$ after the Fourier
transformation\foot{ In this paper we assume $\ap'=\f{1}{2}$ in
most discussions of the type 0 matrix model.} )
\eqn\macnst{\eqalign{ & W^{(-)}_{NS}(E,l)=\int dt\
e^{iEt}e^{-l\Phi^2(t)}, \cr & W^{(-)}_{R}(E,l)=i\int dt\
\f{e^{iEt}}{E}\s{l}~\dot{\Phi}(t)e^{-l\Phi^2(t)},\cr &
W^{(+)}_{NS}(E,l)=\int dt\ e^{iEt} e^{-l\dot{\Phi}^2(t)}, \cr &
W^{(+)}_{R}(E,l)=\int dt\ \f{e^{iEt}}{E}
\s{l}~\Phi(t)e^{-l\dot{\Phi}^2(t)}. \cr}} Notice that
$W^{(-)}_{NS}$ and $W^{(+)}_{R}$ were the same as those proposed
in \TT\ up to a constant factor. One important property of
\macnst\ is the invariance under $Z_2$ action $(-1)^{F_L}$ which
relates the brane $|B(+)\lb$ at $\mu$ to the other brane
$|B(-)\lb$ at $-\mu$ in the $N=1$ Liouville theory. In the fermion
picture of the matrix model, this action is equivalent to the
transformation of fermions into their holes and the replacement of
$x$ with $p=\dot{x}$ at the same time \six. Indeed we can see that
if we replace $\Phi$ with its momentum $\dot{\Phi}$, then we can
get $W^{(+)}_{NS,R}$ from $W^{(-)}_{NS,R}$. Note also that the
expression of $W^{(-)}_{R}$ can be rewritten\foot{ Here we
determined the integration constant by requiring the function $G$
is odd function since we consider the RR-field.}
 by a
partial integration as \eqn\wrrm{ W^{(-)}_{R}(E,l)=\int dt\
e^{iEt}\s{l}\ G(\Phi(t)),\ \ \ (G(x)\equiv \int^x_0 dy e^{-ly^2}).
}

Below we would like to check that the correspondence between the
macroscopic operators and the Euclidean D-brane boundary states
explicitly by computing their one-point functions. In the matrix
model we can diagonalize the matrix $\Phi$ by the gauge symmetry
and regard the eigenvalues as free fermions. We can replace the
trace with a boson $\vp$ via the bosonization $\psi\bar{\psi}\sim
\de \vp$ of the massless Dirac fermion (see e.g.\KlR). Then we get
two bosonic fields which are identified with the spacetime
massless scalar fields $\vp_{NS,R}$ in the NSNS and RR sector.
Following this method, we can obtain the wave functions\foot{This
definition of $F$ is different from \KlR\ by a factor $k$.} (or
one-point functions for fixed $l$) $F^{(\pm)}_{NS,R}(k,l)$  by
\eqn\maci{W^{(\pm)}_{NS,R}(E,l) =\int dk F^{(\pm)}_{NS,R}(k,l)\
\vp(E,k)_{NS,R}.} By using the classical trajectories \eqn\trj{
x(t)=\s{2\mu}\cosh(\tau)\ \ (\mu>0), \ \ \
x(t)=\s{2|\mu|}\sinh(t)\ \ (\mu<0),} we find (see the formula
(A.1) in the appendix) \eqn\formn{\eqalign{& F^{(-)}_{NS}(k,l)
=\f{e^{-l\mu}}{2}k\ K_{ik/2}(l|\mu|), \cr & F^{(+)}_{NS}(k,l)=
\f{e^{l\mu}}{2}k\ K_{ik/2}(l|\mu|)      , \cr &
F^{(-)}_{R}(k,l)=i\f{k}{E}\f{\s{\mu l}}{\s{2}}e^{-\mu
l}(K_{\f{1}{2}+i\f{k}{2}}(l|\mu|)-K_{\f{1}{2}-i\f{k}{2}}(l|\mu|))
, \cr & F^{(+)}_{R}(k,l)=\f{k}{E}\f{e^{l\mu}\s{\mu
l}}{\s{2}}(K_{ik/2+1/2}(l|\mu|) +K_{ik/2-1/2}(l|\mu|)). \cr}} In
actual computations of amplitudes (e.g. the annulus amplitude) we
can pick up poles of propagators and replace $k$ with the energy
$E$. To get the standard $\al=2$ unit, we have to scale as $E\to
2E$. In general the macroscopic operator represents the
expectation value of the massless scalar field $\vp$ in the two
dimensional string theory (see e.g. \GM). Indeed our results
\formn\ are consistent with the wave functions \six\ computed in
the minisuperspace approximation.

As a final step, we would like perform the Laplace
transformation\foot{To be more precise, we should multiply the
extra factor $\s{l}\mu$ only in the RR sector, which comes from
the zeromode insertion of the boundary interaction $\mu_B\int
\eta\psi e^{\phi/2}$.} $\int d\phi\ e^{-\mu_B^2 l}$ from the
function of fixed $l$ to that of the boundary cosmological
constant $\mu_B$ as in the bosonic string case \fzztb. It is also
useful to introduce a new parameter $s$ which is related to
$\mu_B$ as follows \eqn\bcos{ \mu_B^2=2\sinh^2(\pi s)|\mu| \ \ \
(\ep\cdot sign(\mu)<0),\ \ \ \ \ \mu_B^2=2\cosh^2(\pi s)|\mu| \ \
\ (\ep\cdot sign(\mu)>0).} Note that this definition of $s$ is
same as that in boundary Liouville theory after the
renormalization of $\mu_B$ and $\mu$. Finally we obtain the
following results (at $\al=2$) after the Laplace
transformation\foot{Here we have neglected a constant phase factor
like $i$.} using (A.2) in the appendix
 \eqn\laplace{\eqalign{&
\ti{F}^{(+)}_{NS}(E,s)=\ti{F}^{(-)}_{NS}(E,l) =\f{\pi\cos(2\pi
sE)}{\sinh(\pi E)}, \cr & \ti{F}^{(\ep)}_{R}(E,s)=\f{\pi\sin(2\pi
sE)}{\cosh(\pi E)} \ \ \ (\ep\cdot sign(\mu)<0), \cr &
\ti{F}^{(\ep)}_{R}(E,s)=\f{\pi\cos(2\pi sE)}{\cosh(\pi E)} \ \ \
(\ep\cdot sign(\mu)>0).}} Indeed these results agree with the one
point function in $N=1$ boundary Liouville theory \refs{\FuHo, \ARS}\
up to the leg
factors.
 In other words the two point function of two macroscopic loop
operators is the same as the open string one-loop partition
function computed in $N=1$ Liouville theory.

It is also possible to construct macroscopic operators in type 0A
matrix model. This matrix model can be made from $N+q$ D0-branes
and $N$ anti D0-branes \refs{\six, \TT}. The non-zero value of $q$
is proportional to the non-zero RR-flux. This model includes the
complex tachyon field and gauge fields. As noted in \refs{\DKR,
\TT, \GTT}, we can obtain the correct operator\foot{Note that in
this theory there is no propagating field in RR-sector.} by
replacing $e^{-l\Phi(t)^2}$ with $e^{-l\Phi(t)\bar{\Phi}(t)}$ for
$W^{(-)}_{NS}$ in \macnst. We can also define $W^{(+)}_{NS}$ by
the similar $Z_2$ action in 0A model. Since we can reduce the
complex matrix to real eigenvalues (equivalent to the model \JY)
by gauge transformations \refs{\six, \Ka}, the computation can be
done as before. The non-trivial point is that the classical
trajectory is now given by $x(t)^2=\mu+\s{\mu^2+M}\cosh(2t)\ \
(M\equiv q^2-\f{1}{4})$. Then we find the wave functions instead
of \formn \eqn\formnq{F^{(\mp)}_{NS}(k,l) =\f{e^{\mp l\mu}}{2}k\
K_{ik/2}(l\s{\mu^2+M}).} We can also see the same result as
\laplace\ if we define the parameter $s$ by \eqn\paraq{\mu_B^2-\ep
\mu=\s{\mu^2+M}\cosh(2\pi s).} Even though we have no known
comparable results of the boundary states due to the presence of
RR-flux background, our result will give a strong prediction about
it.

\subsec{Possible Relation to Boundary String Field Theory}

Since the macroscopic operators, which are originally operators
in open string theory, make holes on the world-sheet
from the viewpoint of non-critical string, they can also be regarded
as closed string states. Following the general principle of holography in
string theory such as AdS/CFT duality \ads, the operator in open string
theory dual to a closed string field can be determined from the
coupling between closed strings and open strings.

To know the couplings we can employ a certain open string field
theory. We would like to argue that the boundary string field
theory \BSFT\ is most suitable one for our purpose as already
suggested in \TT. Indeed the exponential factor $e^{-l\Phi^2}$
looks like the open string tachyon potential $e^{-T^2}$ in the
boundary superstring field theory \BSFTS. Here we are considering
the formulation in the non-critical superstring. In superstring
the string field action is the same as the disk partition function
\BSFTS. In the presence of the closed string field $\vp(x^0)$, we
can find the shift of string field theory action (i.e. the
couplings to the closed string) $\delta S_{BSFT}$ on multiple
unstable D0-branes as follows \eqn\multiple{ \eqalign{\delta
S_{BSFT}&=\int DX^0 D\psi^0 D\eta DF\ \vp(X^0)\ \Tr\left[
\exp\left(-\int_{\de\Sigma} d\tau d\theta (\Gamma T(X^0)+ \Gamma
D_{\theta} \Gamma )\right)\right], \cr &=\int DX^0 D\psi^0 D\eta \
\vp(X^0)\ \Tr\left[ \exp\left(-\int_{\de\Sigma}
d\tau(T^2+\eta\psi^0\dot{T}+\eta\dot{\eta})\right)\right],\cr}}
where $\Gamma=\eta+\theta F$ is the boundary fermionic superfield
which represents the Chan-Paton degree of freedom as is familiar
in BSFT. Since in the non-critical string the boundary length is
given by $l$, we can identify $\int_{\de\Sigma} d\tau=l$. Also
note the fermionic coordinate $\theta$ scales as $l^{\f{1}{2}}$.
If we look at only zero-modes of \multiple, then we can reproduce
the correct operators $W^{(-)}_{NS}$ and $W^{(-)}_{R}$ in \macnst\
by identifying the tachyon field $T$ with \ the matrix $\Phi$. To
see this result in RR-sector clearly, note that in this case there
exists a fermionic zeromode of $\psi^0$. Also notice that the RR
field $\vp_{R}(x^0)$ (in $(-\f{1}{2},-\f{3}{2})$ picture) should
be proportional to the gauge potential $C \sim \f{i}{k}e^{ikx^0}$
since we always normalize both NSNS massless scalar field and RR
1-form field strength by $e^{ikx^0}$ \TT. The other operators
$W^{(+)}_{NS,R}$ can be obtained from the $Z_2$ transformation.
The operators in 0A theory can also be derived from the boundary
string field theory of brane-antibrane systems \BSFTBA\ in the
same way.

Even though we cannot justify the irrelevance of massive modes of
the field $X^0$, which will lead to higher derivative terms of
$\Phi(t)$, this is not so unnatural since there are no massive
on-shell fields in two dimensional string theory\foot{ This may be
related to the fact that the action of $c=1$ matrix model does not
include any higher derivatives. This might be one of the most
confusing points if we strictly regard the world-volume theory of
unstable D0-branes as the matrix model itself.}. It would be an
interesting future problem to formulate a complete boundary string
field theory for the non-critical string.

\newsec{Putting D-branes in Type 0 Matrix Model}

\subsec{D-branes Localized in Time Direction}

Consider the type 0B model and put a macroscopic operator at time
$t_0$. In the dual two dimensional type 0B theory this means that
there is one Euclidean D-brane localized at time $t_0$.
The brane  extends along the Liouville direction after the Laplace
transformation. First we discuss the operator in the
NSNS-sector. This corresponds to a brane-antibrane system of
the Euclidean D-brane. Then the operator $W^{(-)}_{NS}$ is simply
given by \eqn\opma{\int \f{dl}{l}e^{-l\mu_B^2}e^{-l\Phi^2(t_0)}
=-\log\left(1+\f{\Phi^2(t_0)}{\mu_B^2}\right).} Here we have
determined its constant part such that its value is zero at
$\Phi=0$ and this assumption is consistent with the boundary state
computations as we will see below.
 Thus this is
represented by the following deformation of the action
\eqn\actiond{S=\int dt \Tr\left[
\f{1}{2}\dot{\Phi}^2+\f{1}{2}\Phi^2 +\f{\ap}{2} \delta(t-t_0)
\log\left(1+\f{\Phi^2(t_0)}{\mu_B^2}\right)\right],} where $\ap$
is proportional\foot{Here we believe that the number of D-branes
is neither quantized nor positive since we consider instantonic
D-brane localized in the time direction. We can choose any real
number of $\ap$ as we can assume any coefficient of the
macroscopic operator. On the other hand, the number of unstable
D0-branes is quantized and positive because they are eigenvalues
of the matrix \MV \KMS \MTV.} to the number of D-branes. Below we
assume $\mu>0$ without losing generality due to
the $Z_2$ symmetry $\mu\lr -\mu$.

The equation of motion is given by \eqn\eomr{\f{d^2\Phi(t)}{dt^2}
=\Phi(t)+\ap\delta(t-t_0)\f{\Phi(t_0)}{\Phi^2(t_0)+\mu_B^2}.} To
analyze this background in the matrix model, let us apply a
semiclassical approximation of the system of fermions. The
fermions form a Fermi sea in the phase space $(x,p)$. The Fermi
surface will be deformed from that of the ground state
$p^2-x^2=-2\mu$ due to the delta functional interaction in \actiond. By
integrating \eomr\ we can see that the momentum $p$ is shifted by
$\f{\ap x}{x^2+\mu_B^2}$ at the time $t_0$. Thus at $t=t_0$ we get
the deformed Fermi surface \eqn\fermieom{(p_0-\f{\ap
x_0}{x_0^2+\mu_B^2})^2=x_0^2-2\mu.} The time evolution for $t>t_0$
can be easily obtained from
 \eqn\canon{\eqalign{
 2x_0&=e^{-t}(x+p)+e^{t}(x-p), \cr
2p_0&=e^{-t}(x+p)-e^{t}(x-p).}}

The closed string field corresponds to the fluctuation of the
Fermi surface \DJ \poc \poreview. We can extract the form of
closed string field $\vp_{NS}(t,\phi)$ in the late time asymptotic
region $\phi\to -\infty, t\to +\infty$ by using the identification
\refs{\poc, \poreview} \eqn\fulc{p\sim
x-\f{\mu+\de_{+}\vp(t,\phi)_{NS}}{x},\ \ \ (x=-e^{-\phi}\to
-\infty).} By substituting \canon\ into \fermieom\ and assuming
that $\ap$ is small, we obtain \eqn\flucc{
\de_{+}\vp(t,\phi)_{NS}=\f{\f{\mu^2}{4}e^{2(t+\phi)}-e^{-2(t+\phi)}}
{\f{\mu^2}{4}e^{2(t+\phi)}+e^{-2(t+\phi)}+\mu+\mu_B^2}\ \ap.} This
value changes from $-\ap$ (in the far past) to $\ap$ ( in the far
future) for fixed $\phi$. This can be regarded as the
time-dependent shift of the cosmological constant from $\mu-\ap$
to $\mu+\ap$. The terms with higher powers of $\ap$ will
correspond to higher order in the perturbative expansions of the
string coupling constant.

 We can also
check that an independent computation\foot{To see this, note that
the boundary state is a source to the equation of massless NSNS
fields as $(\de_{t}+p^2)\vp(t,p)=\delta(t-t_0) F_{NS}(p,s)$. This
can be solved as $\vp(t,p)=\f{\sin(pt)}{p} F_{NS}(p,s)$.} by using
the boundary state (see \laplace) leads to the same result up to
the leg factor\foot{Here we put the phase $(\f{\mu}{2})^{ip/2}$
since this is the time delay in the Fermi sea picture as can be
seen from the classical trajectory $x=\s{2\mu}\cosh(t)\sim
\s{\f{\mu}{2}}e^{t}\ \ (t\to\infty)$.} (see the formula (A.3) in
the appendix)
\eqn\bounc{\eqalign{\de_{+}\vp(t,\phi)_{NS}&=\int^{\infty}_{-\infty}
dp \f{\sin(pt)\cos(\pi ps)}{\sinh(\pi
p/2)}(\f{\mu}{2})^{ip/2}e^{ip\phi} \cr &=
\f{2\sinh(2t+2\phi+\log(\mu/2))}{\cosh(2t+2\phi+\log(\mu/2))+\cosh(2\pi
s)}.}} Indeed this expression \bounc\ exactly coincides with
\flucc\ setting the normalization $\ap=2$ (note the relation \bcos
). Another operator $W^{(+)}_{NS}$ can be treated in the same way.
We have only to replace $x$ with $p$ in \fermieom\ etc. We obtain
the same result as \bounc.

Next consider the operator $W^{(-)}_{RR}$ in the RR-sector and
perform the Laplace transformation as before
 \eqn\opmr{\mu_B\int dl e^{-l\mu_B^2} \int^{\Phi}_0 dy e^{-ly^2}
=\mu_B \int^{\Phi}_0 dy \f{1}{y^2+\mu_B^2}
=\arctan\left(\f{\Phi}{\mu_B}\right).} The corresponding deformed
Fermi surface is given by
\eqn\fermieomr{(p_0-\f{\ap\mu_B}{x_0^2+\mu_B^2})^2=x_0^2-2\mu.}
The analysis of the massless scalar field $\vp_{RR}(t,\phi)$ in
the asymptotic region can be done as before and we get the result
using the formula (A.3) in the appendix
 \eqn\bounrr{\de_{+}\vp^{(-)}_{RR}(t,\phi)=
\f{4\sinh(t+\phi+\log(\mu/2))\sinh(\pi
s)}{\cosh(2t+2\phi+\log(\mu/2))+\cosh(2\pi s)}.} We can also find
the result for $W^{(+)}_{RR}$ by replacing $x$ with $p$
\eqn\bounrr{\de_{+}\vp^{(+)}_{RR}(t,\phi)=
\f{4\cosh(t+\phi+\log(\mu/2))\cosh(\pi
s)}{\cosh(2t+2\phi+\log(\mu/2))+\cosh(2\pi s)}.}
Again the matrix model results agree with the world-sheet computations.
The discussions in 0A model can also be done in an analogous way
and will not be discussed in detail here.

Finally let us mention that for large values of $\ap$ (e.g. large
number of D-branes, $\ap >> \f{(\mu+\mu_B^2)\s{\mu}}{\mu_B}$)
the deformed Fermi surface goes beyond the
singular region $p=\pm x$. Then the non-perturbative corrections
become important. However, these backgrounds themselves are
well-defined in type 0B string
 unlike the situation in bosonic
 string. Their qualitative behaviors are rather clear in our
 Fermi sea picture. In this sense the description of D-branes
 by using the matrix model formalism discussed will be a much
stronger method than the usual perturbative formalism of boundary
states. For example, we can see that for a large value of $\ap$
the system includes high energy fermions which may be interpreted
as high energy decaying branes \refs{\GSSB, \RT, \RTS, \KMS}
(`sinh-brane': the second trajectory
of \trj) rather than closed strings.

\subsec{Continuous Array of D-branes}

Next let us consider putting infinitely many D-branes in the time
direction. If we assume the interval is $\delta t$, the source
term is proportional to $\sum_{n}\alpha\delta(t-n\delta t)$.
 We can replace the sum with the integral
$\ti{\alpha} \int dt$ for a finite constant $\ti{\ap}$ by taking
the limit $\delta t \to 0$ at the same time. In other words, we
put a macroscopic operator \macnst\ with $E=0$. Then we get the
matrix quantum mechanics with the deformed potential\foot{ We can
also regard the deformed matrix model as that after a double
scaling limit. Let us start with the matrix quantum mechanics with
the action \eqn\deform{S=\beta \int dt \Tr
[\f{1}{2}\dot{\ti{\Phi}}^2 +\f{1}{2}\ti{\Phi}^2 -c\ti{\Phi}^4+
\f{\ap_0}{2} \log(1+\f{\ti{{\Phi}^2}}{\mu_{B0}^2})],} where $c$ is
a finite constant. Then we can define the double scaling limit (in
the notation of \KlR) as follows \eqn\scale{\beta \sim N\to
\infty,\ \ \ti{\Phi}=\Phi /\s{\beta}, \ \ \beta\ap_0=\ti{\ap}, \ \
\beta\mu_{B0}^2=\mu_{B}^2.} The rescaled quantities $\Phi,\ap$ and
$\mu_{B}^2$ are kept finite. After taking this double scaling
limit, we can neglect the quartic term $c\ti{\Phi}^4$, while the
quadratic and logarithmic terms are relevant. The final model is
the same as what we have discussed above.}
\eqn\deformpo{U(\Phi)=-\Phi^2+\ti{\ap}\log\left(1+\f{\Phi^2}{\mu_B^2}\right).}
for the NSNS operator of 0B model. Since in this section we assume
$\mu$ takes both positive and negative value, we can concentrate
on one of the operators i.e. $W^{(-)}$.

As in the usual $c=1$ matrix model we can diagonalize the matrix
$\Phi$ and represent each of its eigenvalues by $x$.
 At the large value of $|x|$
(i.e. weak coupling region) the original $-x^2$ term is dominant.
In the strongly coupled region, however, the potential is
substantially modified.  For example, if $\mu<0,\ \mu_B^2>>1$ and
$\f{\ap}{\mu_B^2}=$finite$(>1)$, then there is a third Fermi
surface around $x=0$ in addition to the usual left and right ones
(see Fig.1 and Fig.2). The numbers of (semi-stable) bound states
$n_b$ can be estimated by $n_b\sim \mu_B^2>>1$. This part is
expected to describe a new sector hidden inside the Liouville
potential in the strongly coupled region of the two dimensional
string theory. On the other hand, when $\ti{\ap}$ is negative, the
physics will be qualitatively similar to the usual case. \fig{The
shape of the deformed potential $U(x)$ at $\ti{\ap}=300,\mu_B=10$.
The horizontal axis and vertical line represents the values of $x$
and $U(x)$.} {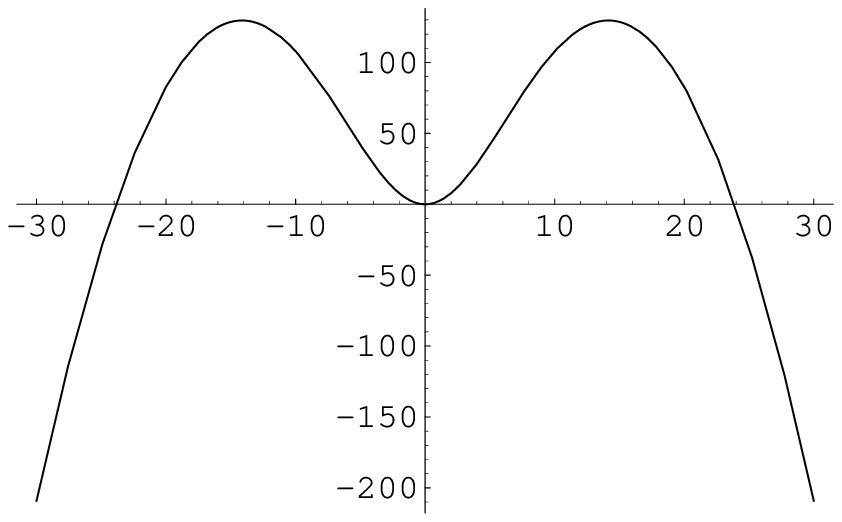 }{2.0truein} \fig{The structure of the
Fermi surface for various values of $\mu$ at
$\ti{\ap}=300,\mu_B=10$. We have shown the contours for various
fixed values of $\mu$. We can see there are three isolated Fermi
surfaces for appropriate values of $\mu$. The horizontal axis and
vertical line represents the values of $x$ and
$p$.}{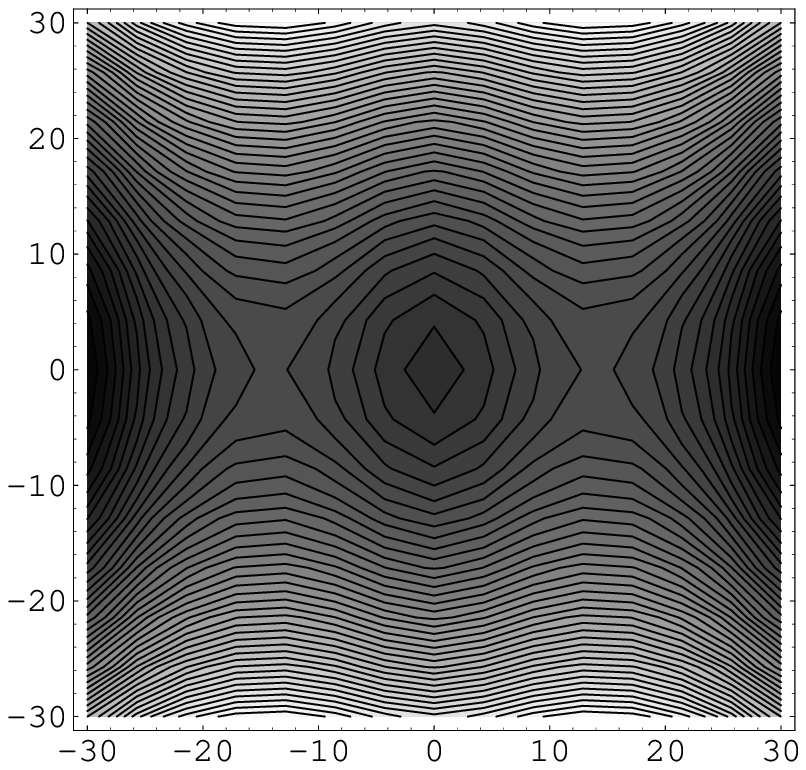}{2.0truein}

For this NSNS-sector operator it might also be possible to assume
$\mu_B^2<0$. When $\ti{\ap}$ is positive, we have a third Fermi
surface at $-|\mu_B|<x<|\mu_B|$, which has no bottom of potential.
In this case we may have a non-perturbative instability.
 When $\ti{\ap}$ is
negative, there is a large wall at $x=\pm\mu_B$ and there is no
non-perturbative mixing (D-instanton effect \ponon) between the left and
right Fermi sea.

We can also analyze the condense of D-branes in the 0A model. In
this case the potential becomes $U(x)=-x^2+\f{q^2-\f{1}{4}}{x^2}+
\ti{\ap}\log\left(1+\f{x^2}{\mu_B^2}\right)$. Since the
qualitative feature of the results will be similar to the previous cases,
we will not discuss this in detail.

On the other hand, the macroscopic operator in the
RR-sector of 0B theory leads to the deformed potential
\eqn\deformporr{U(\Phi)=-\Phi^2+\ti{\ap}\arctan\left(\f{\Phi}{\mu_B}\right).}
We can again study its property in the same way as before.
 For a usual (`BPS-like') D-brane
the deformation of the potential is given by the sum of the
contributions in NSNS
\deformpo\ and RR
\deformporr\ sector.

Finally we would like to compare our results in type 0 string with
those of bosonic string. In the bosonic string the deformed
potential is given by $U(\Phi)=-\Phi^2+\ti{\ap}\log(\Phi+\mu_B)$.
The model is problematic when $x+\mu_B\leq 0$. This is due to the
fact that in the bosonic string case we consider only one Fermi
sea $x>0$ and that we must assume $\mu_B$ is positive. This is in
contrast with the type 0 case where there is no singularity as is
obvious from \deformpo \deformporr.

\newsec{Conclusion and Discussion}

In this paper we constructed a complete set of macroscopic
operators in type 0 matrix model. This gives a realization of
Euclidean D-branes (FZZT-brane) in the two dimensional type 0
string theory. We checked that the operators correctly reproduce
the one-point functions of boundary states in the super Liouville
theory. We have pointed out their possible relation to the boundary
superstring field theory.
We also showed how to represent the presences of such
D-branes in the semiclassical picture of the phase space. They are
given by specific deformations of Fermi surfaces. Finally we
considered putting a continuous array of the D-branes. This leads
to a matrix model with a deformed potential. In some cases we found
that it has an additional Fermi surface, which may be interpreted
as a new sector hidden inside the strongly coupled region. This
may give an important hint as to blackholes
 in two dimension (e.g.\wittenbh \KKK) since the region inside the
horizon is typically strongly coupled. Indeed we can compute how
the black hole mass operator $\de X\bar{\de} X e^{2\phi}$ will be
induced in the presence\foot{In the presence of unstable D0-branes
an interesting result on the backreaction to discrete states was
obtained quite recently in \SenR.} of the D-branes discussed in
section 3.2. We can find that it is proportional to $e^{2\pi s}$
for large $s$ by using the one-point function \laplace. This seems
to be consistent with the previous result that the new sector
appears when $\mu_B^2$ is very large.

In these examples we can describe the backgrounds with D-branes
including non-perturbative corrections in the type 0 matrix model.
This will be a good toy model when we consider solving non-linear
backreactions of D-branes non-perturbatively, which is usually
very difficult in ten dimensional superstring.

\vskip 0.4in

\centerline{\bf Acknowledgments}

The author is grateful to
D.Gaiotto, S. Gukov, A. Jevicki,  F. Larsen, S. Minwalla,
 N. Seiberg, A. Strominger, S. Terashima, N. Toumbas and S. Yamaguchi
for useful discussions. The work was supported in part by DOE
grant DE-FG02-91ER40654.

\appendix{A}{Useful Identities}

In section 2.1 we have used the following identities (see also \Ma
\six \TT) \eqn\formulaw{\eqalign{& K_{iE}(\mu
l)=\int^{\infty}_{0}ds\ e^{-\mu l\cosh s}\cos(Es), \cr &
K_{\f{1}{2}+iE}(\mu l)+K_{\f{1}{2}-iE}(\mu l)
=\int^{\infty}_{0}ds\ e^{-\mu l\cosh s}\cosh (s/2)\cos(Es), \cr &
K_{\f{1}{2}+iE}(\mu l)-K_{\f{1}{2}-iE}(\mu l)
=i\int^{\infty}_{0}ds\ e^{-\mu l\cosh s}\sinh (s/2)\sin(Es), }}
and \eqn\formulao{\eqalign{& \int^{\infty}_{0}\f{dl}{l}e^{-\mu l
\cosh(2\pi s)}K_{iE}(l\mu) =\f{\pi\cos (2\pi sE)}{E\sinh (\pi E)}
, \cr & \int^{\infty}_{0}\f{dl}{l}(l\mu) e^{-\mu l \cosh(2\pi
s)}\cosh (\pi s)(K_{iE+1/2}(l\mu)+K_{iE-1/2}(l\mu)) =\f{\pi\cos
(2\pi sE)}{\cosh(\pi E)}, \cr & -i\int^{\infty}_{0}\f{dl}{l}(l\mu)
e^{-\mu l \cosh(2\pi s)}\sinh (\pi
s)(K_{iE+1/2}(l\mu)-K_{iE-1/2}(l\mu)) =\f{\pi\sin (2\pi sE)}{\cosh
(\pi E)}.}} In section 3.1 we have employed
 \eqn\formcc{\eqalign{ \int^{\infty}_{-\infty}\f{\cosh
ax}{\sinh \pi x} e^{ixy}&=\f{i\sinh y}{\cosh y+\cos a} \cr
\int^{\infty}_{-\infty}\f{\sinh ax}{\cosh \pi x}
e^{ixy}&=\f{2i\sin a/2 \sinh y/2}{\cosh y+\cos a} \cr
\int^{\infty}_{-\infty}\f{\cosh ax}{\cosh \pi x} e^{ixy}&=\f{2\cos
a/2 \cosh y/2}{\cosh y+\cos a} .}}
\listrefs

\end